\documentclass[convention,peer-reviewed]{aesconf}

\graphicspath{{./}{figures/}}

\usepackage[utf8]{inputenc}

\usepackage{microtype}

\usepackage[numbers,square]{natbib}
\usepackage{mathtools}
\usepackage[refpage]{nomencl}
\usepackage{booktabs}
\usepackage{color}
\usepackage{xurl}
\usepackage[refpage]{nomencl}
\usepackage[hidelinks]{hyperref}
\usepackage{xspace}
\usepackage{tablefootnote}
\usepackage{gensymb}

\title{Universal Spatial Audio Transcoder}

\author[2,*]{Amaia Sagasti}
\author[1]{Davide Scaini}
\author[1,2]{Daniel Arteaga}

\affil[1]{Dolby Laboratories.}
\affil[2]{Universitat Pompeu Fabra, Barcelona, Spain.}
\affil[*]{\emph{Work done as part of the author's internship at} Dolby Laboratories.}

\newcommand{\Dmat}{\ensuremath{\mathbf{D}}\xspace}
\newcommand{\Tmat}{\ensuremath{\mathbf{T}}\xspace}
\newcommand{\Gmat}{\ensuremath{\mathbf{G}}\xspace}
\newcommand{\Smat}{\ensuremath{\mathbf{S}}\xspace}
\newcommand{\Dspk}{\ensuremath{\mathbf{D}_\text{spk}}\xspace}

\correspondence{Daniel Arteaga}{daniel.arteaga@dolby.com}

\lastnames{Sagasti, Scaini, and Arteaga}

\shorttitle{Universal Spatial Audio Transcoder}

\renewpagestyle{plain}{
	\sethead[][][] %
			{}{}{} %
	\setfoot[][\thepage][] %
			{}{\thepage}{} %
}
\begin{document}

\twocolumn[
\maketitle %
\begin{onecolabstract}
This paper addresses the challenges associated with both the conversion between different spatial audio formats and the decoding of a spatial audio format to a specific loudspeaker layout. Existing approaches often rely on layout remapping tools, which may not guarantee optimal conversion from a psychoacoustic perspective. To overcome these challenges, we present the \emph{Universal Spatial Audio Transcoder} (USAT) method and its corresponding open source implementation. USAT generates an optimal decoder or transcoder for any input spatial audio format, adapting it to any output format or 2D/3D loudspeaker configuration. Drawing upon optimization techniques based on psychoacoustic principles, the algorithm maximizes the preservation of spatial information. We present examples of the decoding and transcoding of several audio formats, and show that USAT approach is advantageous compared to the most common methods in the field. 
\end{onecolabstract}
]

\section{Introduction}

Various formats exist for representing spatial audio, ranging from layout-independent approaches, such as Ambisonics or object-based approaches, to layout-specific coding formats, such as traditional multichannel configurations like 5.1 or 7.1.4. While layout-independent formats offer the advantage of independence from the specific speaker arrangement, they necessitate a dedicated decoder for accurate reproduction. Conversely, formats tailored to particular loudspeaker layouts do not need specific decoders when reproduced over the ideal setup.  However, practical scenarios frequently differ from the intended playback setup, necessitating adaptation to preserve spatial information and maintain the overall listening quality. In addition to these considerations, the diverse landscape of spatial audio formats often requires transcoding between them. All these aspects together underscore the importance of having decoding and transcoding tools.

It is important to recognize that spatial audio decoding can be viewed as a particular case of transcoding, wherein the output spatial audio format is defined by a multichannel mix tailored for the actual loudspeaker configuration. Throughout this paper, when we will mention transcoding, we will encompass both the proper transcoding from one spatial audio format to another (e.g., from 5.1 to Ambisonics) and the decoding of a specific spatial audio format to match a particular loudspeaker layout (e.g., decoding 7.1.4 to fit an irregular 5.1 setup).

Various approaches have been employed to tackle the transcoding problem. Despite their differences, they typically treat the input spatial audio format as a collection of virtual point sources for the desired output spatial audio format. Frequently, these approaches leverage a panning law for the conversion process. For instance, the widely adopted Ambisonics decoding method AllRad  \cite{zotter_allrad_2012} involves decoding to an intermediate  loudspeaker setup consisting of a regular layout of virtual loudspeakers, followed by applying vector-base amplitude panning (VBAP) \cite{pulkki_vbap_97} to remap the virtual layout to the real loudspeaker layout. When decoding multichannel configurations such as 5.1 or 7.1.4 into non-regular loudspeaker layouts, popular options include mapping to the closest loudspeaker equivalents or remapping the intended layout into the real layout using a panning law \cite{ando_conversion_2010, schmele_remapping_2013}. Layout remapping is also commonly chosen for conversion between different multichannel formats \cite{schmele_remapping_2013}, often supplemented by ad-hoc rules to enhance the process, specially when downmixing \cite{dolby_downmix}.  Finally, transcoding a multichannel format to Ambisonics is often accomplished by treating each channel of the mix as a separate virtual point source and encoding them in Ambisonics \cite{waves}. 

It is to be noted that the existing approaches relying on layout remapping tools may not necessarily guarantee optimal conversion or decoding from a psychoacoustic perspective. In addressing the latter concern, the IDOHA decoder \cite{scaini_ambisonics_2014} proposed an alternative Ambisonics decoding method that achieves optimal decoding through the optimization of a psychoacoustics-based cost function \cite{arteaga_ambisonics_2013}. The IDHOA decoder was also adapted to decode a wavelet-based spatial audio format to non-regular layouts \cite{scaini_wavelet_2020}. However, the principles behind the IDHOA decoder are not limited to Ambisonics or wavelet-based spatial audio, but in fact they can be extended to any channel-based linear-encoding spatial audio format.

Drawing upon the optimization techniques that formed the foundation of IDHOA, this paper introduces the \emph{Universal Spatial Audio Transcoder} (USAT) algorithm. Based on the minimization of a perceptually motivated cost function, the USAT algorithm is designed to generate an optimal transcoder or decoder to adapt the input to any output spatial audio format or 2D/3D loudspeaker configuration. We also provide an open-source implementation of the algorithm in Python \cite{usat_github_2024}.

This paper is structured as follows. Section \ref{sec:methods} provides a detailed description of the USAT algorithm. In Section \ref{sec:applications}, we explore multiple applications of USAT and compare its performance with existing methods. Our main findings and their implications are discussed in Section \ref{sec:conclusions}. In the Appendix we summarize the notation.

\section{Algorithm description} \label{sec:methods}

\begin{figure*}
    \centering
    \includegraphics[width=\textwidth]{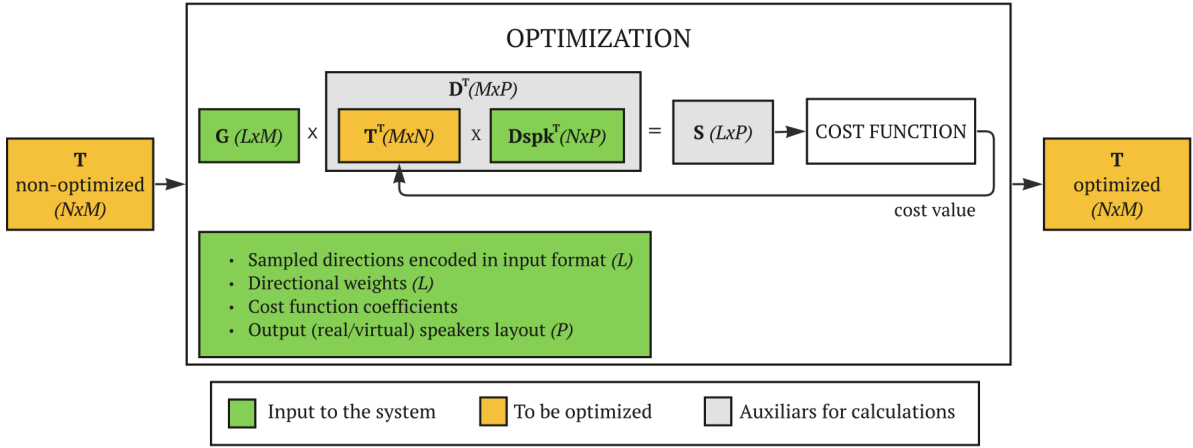}
    \caption{Overview of the optimization process in USAT.
    Dimensions $M$ and $N$ indicate the number of input and output channels, respectively; $L$, the number of sampled directions, and $P$ the number of loudspeakers in the real or virtual layout. 
    }
    \label{fig:system-schema}
\end{figure*}

\subsection{Overview of the algorithm}\label{cap:complete-process-section}

The algorithm proceeds in three basic steps (see also Fig.~\ref{fig:system-schema}):
\begin{enumerate}
    \item \emph{Encoding, transcoding and decoding matrices setup.} The pertinent matrices for the problem are either computed or initialized.
    
        The \emph{encoding matrix} \Gmat describes how a set of virtual sources located at directions sampled across the sphere (or the circle in 2D) are encoded into the input audio format. It serves as an input to the problem. The encoding matrix, in conjunction with the set of sampling directions, fully characterizes the input audio format.
        
        The \emph{transcoding matrix} \Tmat maps the input audio format to the output audio format. This matrix is unknown, and it is initialized by either an educated or a random guess.
        
        The \emph{decoding-to-speaker matrix} \Dspk describes how the channels of the output audio format are decoded to a layout of loudspeakers, which can be real or virtual (see below). It serves as another input to the problem. The decoding-to-speaker matrix, along with the loudspeaker layout, fully characterizes the output audio format. %

        The software implementation offers helper tools to compute the encoding and decoding-to-speaker matrices \Gmat and \Dspk for common formats like VBAP and Ambisonics.

    \item \emph{Cost function setup.} Using the foundation of the above-defined matrices, a psychoacoustic cost function is established based on the same principles as IDHOA \cite{arteaga_ambisonics_2013, scaini_ambisonics_2014}, which are rooted on the Gerzon localization fundamentals \cite{gerzon_metatheory_1992} and have had experimental validation \cite{zotter_ambisonics_2019}. There are two primary versions of the cost function. The first one assumes coherence and relies on squared linear terms in the cost function;  the second one assumes incoherence in the decoding and is based on squared quadratic terms in the cost function. %

    \item \emph{Cost function minimization.} The cost function is minimized with respect to the transcoding matrix. The optimized transcoding matrix  \Tmat is  the main outcome of the cost function minimization.
    
\end{enumerate}

\subsection{Encoding, transcoding and decoding matrices}

The system aims to find an optimal \(N\times M\)  transcoding matrix \Tmat for converting an \(M\)-channel input spatial audio format into an \(N\)-channel output format.

The input format is characterized by how a set of virtual sources, positioned at \(L\) directions sampling the sphere, is encoded. Specifically, it is defined by the \(L \times M\) encoding matrix \Gmat, composed by the set of gains \(\{g_{\ell m}\}\), along with the angular locations of these virtual sources \(\{\hat{v}_\ell\}\). These gains encode virtual sources from each of the \(L\) directions into each of the \(M\) channels, e.g. VBAP panning coefficients for each one of the sampled directions into the 7 channels of a 5.0.2 layout.

The output format is defined by its decoding to loudspeakers. Specifically, assuming a loudspeaker layout with \(P\) loudspeakers, the output format is  characterized by the  decoding-to-speaker matrix \Dspk, a \(P \times N\) matrix mapping each of the \(N\) output channels into each of the \(P\) loudspeakers. The characterization of the output format is completed with the angular locations of the loudspeakers \(\{\hat{u}_p\}\). In proper transcoding scenarios, this loudspeaker layout normally represents a virtual loudspeaker arrangement well suited for the output format (e.g., for 1st order Ambisonics output, the matrix could be the basic decoding to a 6-loudspeaker octahedral layout).
 Conversely, in decoding situations, the loudspeaker layout matches a real venue, the number of loudspeakers corresponds to the number of output channels, \(P=N\), and the the decoding-to-speaker matrix is simply an identity matrix, \(\Dspk = \mathbf 1_{N\times N}\).

The \(P \times M\) \emph{decoding matrix} \Dmat is the product of the decoding-to-speakers matrix \Dspk and the transcoding matrix:
\begin{equation}
    \Dmat = \Dspk \, \Tmat
\end{equation}
The decoding matrix \Dmat decodes each one of the input channels into each one of the loudspeakers in the real or virtual layout. 

We call the product of the gain matrix by the transposed decoding matrix the \emph{speaker matrix} \Smat:
\begin{equation} \label{eq:speaker}
    \Smat =  (\Dmat \, \Gmat^T)^T =  \Gmat\, \Dmat^T = \Gmat\,\Tmat^T\, \Dspk^T
\end{equation}
The \(L\times P\) speaker matrix \Smat characterizes the gain coefficients for each one of the \(P\) loudspeakers given the virtual sound source located at each one of the \(L\) encoding directions.\footnote{The appearance of the matrix transpose in Eq.~\eqref{eq:speaker} is a result of the specific ordering of dimensions in the system and gain matrices. This ordering is arbitrary and lacks intrinsic significance. The chosen arrangements are simply a means to closely match the associated open-source code.} In other words, the component \(s_{\ell p}\) of the speaker matrix represents the signal fed to the loudspeaker $p$ while reproducing a virtual signal of unit amplitude coming from direction $\hat v_\ell$. These components will be the building blocks of the cost function.

\subsection{Cost function}

As already mentioned, there are two main components of the cost function: one assuming coherent behaviour in the decoding, and another one assuming incoherent behaviour. In the following, we describe the terms corresponding to both versions, as well as the additional terms that can be added to customize the behaviour.

\subsubsection{Linear decoding terms (coherence)}

In ideal situations, and in real situations at low frequencies, the signal received by the listener is the coherent addition of the signal coming from each one of the loudspeakers. 

Under this coherence hypothesis, the normalized pressure at the listener's position can be taken to be:
\begin{equation}
    P_\ell =\sum_{p=1}^{P} s_{\ell p},
\end{equation}
where $\{s_{\ell p}\}$ are the components of the speaker matrix, see Eq.~\eqref{eq:speaker}. 
Similarly, the normalized acoustic velocity can be taken to be:
\begin{equation}    
    \vec V_\ell =\frac{1}{P_\ell} \sum_{p=1}^{P} s_{\ell p} \, \hat{u}_p
\end{equation}
The normalized velocity vector \(\vec V_\ell\) can be projected in its radial and transverse part as follows \cite{arteaga_ambisonics_2013}:
\begin{subequations}
    \begin{align}
        V^R_\ell &= \vec V_\ell\cdot \hat{v}_\ell = \frac{1}{P_\ell}\sum_{p=1}^{P} s_{\ell p} \, \hat{u}_p\cdot \hat{v}_\ell \\
        V^T_\ell &= \left\lVert{\vec V_\ell\times\hat{v}_\ell}\right\rVert = \frac{1}{P_\ell} \sum_{p=1}^{P} s_{\ell p} \left\lVert\hat{u}_p\times \hat{v}_\ell\right\rVert
    \end{align}
\end{subequations}
The radial part \(V^R_\ell\) represents the desired component of the velocity vector whereas the transverse part, \(V^T_\ell\), represents the unwanted component. In an ideal system \(P_\ell = 1\), \(V^R_\ell = 1\) and \(V^T_\ell = 0\).

From these, three different cost function terms can be defined:
\begin{subequations}
    \begin{align}
        C_P &= \frac{1}{L} \sum_{\ell=1}^{L} (1-P_\ell)^2\, w_\ell  \\
        C_{VR} &= \frac{1}{L} \sum_{\ell=1}^{L} (1-V^R_\ell)^2\, w_\ell \\
        C_{VT} &= \frac{1}{L} \sum_{\ell=1}^{L} (V^T_\ell)^2\, w_\ell
    \end{align}
\end{subequations}
The weighting factor \(w_\ell\) is an optional biasing factor which allows to improve the decoding performance in some regions of the space (at the expense of other regions). A non-biased decoding is given by \(w_\ell = 1\).

Under the coherence hypothesis, these contributions can be interpreted as follows: \(C_P\) is the mean quadratic deviation from the correct pressure level; \(C_{RV}\) is the mean quadratic deviation from the optimal directivity; and, finally, \(C_{TV}\) is the mean quadratic value of the undesired component of the direction.

\subsubsection{Quadratic decoding terms (incoherence)}
In real situations at mid and high frequencies, or far from the sweet spot, the signal received by the listener is normally  better approximated by the incoherent addition of the signal coming from each one of the loudspeakers. 

Under this incoherence hypothesis, the acoustic energy can be approximated by:
\begin{equation}
    E_\ell =\sum_{p=1}^{P} \lvert{s_{\ell p}}\rvert^2
\end{equation}
and the normalized acoustic intensity can be estimated by the so-called Gerzon energy vector:
\begin{equation}
    \vec I_\ell =\frac{1}{E_\ell} \sum_{p=1}^{P} \lvert{s_{\ell p}}\rvert^2 \hat{u}_p
\end{equation}
The vector \(\vec I_\ell\) can be similarly projected into the radial and transverse part as follows:
\begin{subequations}
    \begin{align}
        I^R_\ell &= \vec I_\ell\cdot \hat{v}_\ell = \frac{1}{E_\ell} \sum_{p=1}^{P} \lvert{s_{\ell p}}\rvert^2 \hat{u}_p\cdot \hat{v}_\ell, \\
        I^T_\ell &= \left\lVert{\vec I_\ell\times\hat{v}_\ell}\right\rVert = \frac{1}{E_\ell} \sum_{n=1}^{P} \lvert{s_{\ell p}}\rvert^2 \left\lVert\hat{u}_p\times \hat{v}_\ell\right\rVert.
\end{align}
\end{subequations}

The radial part \(I^R_\ell\) represents the desired component of the intensity vector whereas the transverse part, \(I^T_\ell\), represents the unwanted component. In an ideal system \(E_\ell = 1\), \(I^R_\ell = 1\) and \(I^T_\ell = 0\).

In fact, under the incoherence hypothesis, the value of $E$ correlates with the perceived level, and the radial and transverse intensities are related to the apparent source width (ASW) \cite{zotter_ambisonics_2019} and error deviation ($\delta$):
\begin{subequations} \label{eqs:aswdelta}
    \begin{align}
        \mathrm{ASW} &= \frac{3}{4} \arccos {\lVert \vec I \rVert} =\frac{3}{4} \arccos {\sqrt{(I^R)^2 + (I^T)^2}} 
        \label{eq:asw} \\
        \delta &= \arctan{\frac{I^T}{I^R}} \label{eq:delta}
    \end{align}
\end{subequations}
In an ideal system, for a virtual source $\mathrm{ASW} = \delta = 0$.

Based on the energy and radial and transverse intensity three different cost functions can be defined:
\begin{subequations}
    \begin{align}
        C_E &= \frac{1}{L} \sum_{\ell=1}^{L} (1-E_\ell)^2 w_\ell \\
        C_{IR} &= \frac{1}{L} \sum_{\ell=1}^{L} (1-I^R_\ell)^2 w_\ell \\
        C_{IT} &= \frac{1}{L} \sum_{\ell=1}^{L} (I^T_\ell)^2 w_\ell 
\end{align}
\end{subequations}
Under the incoherence hypothesis, these contributions can be interpreted as follows: \(C_E\) is the mean quadratic deviation from the correct level reconstruction; \(C_{IR}\) is the mean quadratic deviation from the optimal directivity;  and finally, \(C_{IT}\) is the mean quadratic value of the undesired component of the direction.\footnote{Instead of focusing on optimizing $I^R$ and $I^T$, an alternative approach could involve optimizing ASW and $\delta$ instead. However, the main rationale for prioritizing the optimization of $I^R$ and $I^T$ lies in their orthogonality, which ensures independent adjustments to each component of the intensity vector. In contrast, ASW and $\delta$ exhibit a significant negative correlation (decreasing ASW will generally increase $\delta$), making them less conducive to independent optimization efforts.}

\subsubsection{Other cost function terms}

In addition to the psychoacoustic terms above, other terms, playing the role of soft constraints, can be introduced to help the solution achieve desired properties.

It is possible to penalize out-of-phase (i.e. favour in-phase) decoding with an extra cost function term. Two versions are proposed, the linear and the quadratic:
\begin{subequations}
   \begin{align}
        \Phi^\text{lin}_\ell &= \frac{1}{\lvert P_\ell \rvert} \sum_{p=1}^{P} \lvert{s_{\ell p}}\rvert \,\theta(-s_{\ell p}) \\
        \Phi^\text{quad}_\ell &= \frac{1}{E_\ell} \sum_{p=1}^{P} \lvert{s_{\ell p}}\rvert^2\, \theta(-s_{\ell p})
    \end{align} 
\end{subequations}
where $\theta$ is the Heaviside step function. 
From these quantities, the following two cost function terms can be defined:
\begin{subequations}
   \begin{align}
        C_\Phi^\text{lin} &= \frac{1}{L} \sum_{\ell=1}^{L} \left(\Phi^\text{lin}_\ell\right)^2 \,  w_\ell\\
        C_\Phi^\text{quad}  &= \frac{1}{L} \sum_{\ell=1}^{L} \left(\Phi^\text{quad}_\ell\right)^2 \, w_\ell
    \end{align} 
\end{subequations}

Additionally, it is possible to add a symmetry penalty to encourage left-right symmetry in the generated transcoding matrix. Once identified the left-right symmetric pairs in the destination layout, it is possible to quantify the amount of asymmetry present in the decoding (again, both in quadratic and linear versions):
\begin{subequations}
   \begin{align}
        \Delta^\text{lin}_\ell &=\frac{1}{\lvert P_\ell \rvert} \sum_{\text{symmetry pairs }(p,p') } \lvert{s_{\ell p} - s_{\ell p'}}\rvert \\
        \Delta^\text{lin}_\ell &= \frac{1}{E_\ell} \sum_{\text{symmetry pairs }(p,p') } \lvert{s_{\ell p} - s_{\ell p'}}\rvert^2 
    \end{align} 
\end{subequations}
From these quantities, the following two cost function terms can be defined:
\begin{subequations}
   \begin{align}
        C_\Delta^\text{lin} &=  \frac{1}{L} \sum_{\ell=1}^{L} \left(\Delta^\text{lin}_\ell \right)^2 \,w_\ell ,\\
        C_\Delta^\text{quad}  &= \frac{1}{L} \sum_{\ell=1}^{L} \left(\Delta^\text{quad}_\ell \right)^2\, w_\ell.
    \end{align} 
\end{subequations}

Furthermore, a limitation to the total set of gains of the decoding matrix is introduced through another cost function term. By these means, a limitation of the total boost pressure is achieved (e.g., 3 dB): 
\begin{subequations}
   \begin{align}
        \Sigma^\text{lin} &= \frac{1}{NM} \sum_{n=1}^{N} \sum_{m=1}^{M} d_{nm}\, \theta(d_{nm} - d_\text{max}),\\
        \Sigma^\text{quad}  &= \frac{1}{NM} \sum_{n=1}^{N} \sum_{m=1}^{M} (d_{nm})^2\, \theta(d_{nm} - d_\text{max}),
    \end{align} 
\end{subequations}
where $d_\text{max} = 10^{L_\text{max}/20}$, with $L_\text{max}$ being the maximum boost allowed in dB.
Similarly to the cases above, this total gains term has two versions, the linear and the quadratic:
\begin{subequations}
   \begin{align}
        C_\Sigma^\text{lin} &= \frac{1}{L} \sum_{\ell=1}^{L} \left(\Delta^\text{lin}_\ell \right)^2 \,w_\ell, \\
        C_\Sigma^\text{quad}  &= \frac{1}{L} \sum_{\ell=1}^{L} \left(\Delta^\text{quad}_\ell \right)^2\, w_\ell.
    \end{align} 
\end{subequations}

Finally, in some occasions it may be beneficial to enhance the sparsity of the results. One possible way to quantify the non-sparsity of the solution is based on the difference of the L1 and L2 norms of the rows of the speaker matrix:
\begin{subequations}
   \begin{align}
        S^\text{lin}_\ell &= \frac{1}{\lvert P_\ell \rvert} \left[\sum_{p=1}^{P} \lvert{s_{\ell p}}\rvert 
        -
        \left(\sum_{p=1}^{P} \lvert{s_{\ell p}}\rvert^2 \right)^{1/2}\right],
        \\
        S^\text{quad}_\ell &= \frac{1}{E_\ell} \left[ \left(\sum_{p=1}^{P} \lvert{s_{\ell p}}\rvert \right)^2
        -
        {\sum_{p=1}^{P} \lvert{s_{\ell p}}\rvert^2 } \right].
    \end{align} 
\end{subequations}
The following sparsity-enhancing cost function terms can be defined:
\begin{subequations}
   \begin{align}
        C_S^\text{lin} &= \frac{1}{L} \sum_{\ell=1}^{L} \left(S^\text{lin}_\ell \right)^2 \,w_\ell, \\
        C_S^\text{quad}  &= \frac{1}{L} \sum_{\ell=1}^{L} \left(S^\text{quad}_\ell \right)^2\, w_\ell.
    \end{align} 
\end{subequations}

\subsubsection{Total cost function}

The total cost function is formed by the addition of the cost function terms with the corresponding prefactors (denoted below by $c_x$):
\begin{equation}
    \begin{split}
        C &=  c_P\, C_P + c_{VR}\, C_{VR} + c_{VT}\, C_{VT} + c_\Phi^\text{lin}\, C_\Phi^\text{lin} \\
            &\mathrel{\phantom{=}}  + c_\Delta^\text{lin}\, C_\Delta^\text{lin}  + c_\Sigma^\text{lin} C_\Sigma^\text{lin} + c_S^\text{lin} C_S^\text{lin} \\
            &\mathrel{\phantom{=}} + c_E\, C_E + c_{IR}\, C_{IR} + c_{IT}\, C_{IT} + c_\Phi^\text{quad}\, C_\Phi^\text{quad}  \\
            &\mathrel{\phantom{=}}  + c_\Delta^\text{quad}\, C_\Delta^\text{quad}  + c_\Sigma^\text{quad} C_\Sigma^\text{quad}
            + c_S^\text{quad} C_S^\text{quad}.
    \end{split}    \label{eq:cost}
\end{equation} 
The values of the prefactors can be selected at will. To ensure that all minimization terms will scale in a similar way during the optimization process, often linear and quadratic terms will not be mixed together.

\subsection{Cost function minimization}
The optimizer minimizes the cost function $C$ with respect to the transcoding matrix \Tmat, see Eq.~\eqref{eq:cost}. The specific optimizer used in the open-source implementation of USAT is the BFGS optimization method available in the SciPy \cite{scipy} package. Gradients are computed leveraging the automatic differentiation functionality provided by Jax \cite{jax}.

The optimization of the cost function delivers the optimal transcoding matrix \Tmat.

\section{Example applications} \label{sec:applications}

Four different example applications of USAT are presented, demonstrating the diverse range of possibilities offered by the algorithm. The initial example involves decoding 5th order Ambisonics (5OA) to a 7.0.4 multichannel layout. Following this, the second example illustrates the inverse process: transcoding a 7.0.4 multichannel input into 5OA. The third scenario pertains to another decoding case: converting a 5.0.2 multichannel layout to an irregular speaker configuration (3.0.1). Finally, the last example shows the decoding of an arbitrary virtual sound source to a 5.0 layout, making it applicable to audio object decoding.

\begin{table} [tbh]
    \caption{Information about the four example applications of USAT, including  cost function coefficients (only non-zero terms are shown) and optimization time (MacBook Pro M3). See the main text for detailed explanations.}
  \label{fig:table-configurations}
  \small
  \centering
  \begin{tabular}{lcccc}
    \toprule
    Example & 1 & 2 & 3 & 4 \\
    \midrule
    Type & Dec. & Trans. & Dec. & Dec. \\
    Approach & Incoh. & Coh. & Incoh. & Incoh. \\
    \midrule
    Input  & 5OA & 7.0.4  & 5.0.2  & Objects \\
    $M$ & 36 & 11 & 7 & \# obj.\tablefootnote{Alternative, the decoding coefficients could be precomputed for a set of grid positions and then interpolated. See main text.} \\
    Output   & 7.0.4  & 5OA  & 3.0.1 irr. & 5.0  \\
    $N$ & 11 & 36 & 4 & 5 \\
        \midrule
    $c_P$ & - & 5 & - & - \\
    $c_{VR}$ & - & 2 & - & - \\
    $c_{VT}$ & - & 1 & - & - \\
    \midrule
    $c_E$ & 5 & - & 5 & 5 \\
    $c_{IR}$ & 2 & 0.2 & 2 & 2 \\
    $c_{IT}$ & 1 & 0.1 & 1 & 1 \\
    \midrule
    $c_\phi^\text{quad}$ & 10 & - & $10^4$ & $10^4$ \\
    $c_\Delta^\text{quad}$ & 2 & - & - & 2 \\
    $c_S^\text{lin}$ & - & - & $10^{-3}$ & - \\
    $c_S^\text{quad}$ & - & - & $10^{-2}$ & - \\
    \midrule
    Opt. time (s) & 14.6 & 4.3 & 1 & 3.9 \\
    \bottomrule
  \end{tabular}

\end{table}

Information on the basic settings and cost function coefficients is reported in Table~\ref{fig:table-configurations}. It should be noted that we have tried to avoid fine-tuning each one of the examples. Instead, whenever possible we use a common set of coefficients for all the examples. 

In each example, USAT is compared using objective metrics against alternative decoding/transcoding methods. For each example and each transcoding/decoding method, virtual sources are evaluated regarding: (i) level in dB, quantified by $P$ or $E$ depending on the decoding assumption (respectively, coherence or incoherence); (ii) apparent source width (ASW) in degrees [Eq.~\eqref{eq:asw}], and (iii) angular error ($\delta$) in degrees [Eq.~\eqref{eq:delta}].

\subsection{5th order Ambisonics decoding to 7.0.4}

\begin{figure} [tbh]
    \centering 
    \includegraphics{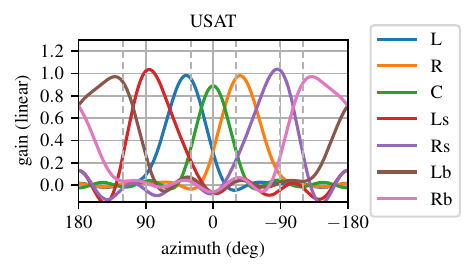} \\
    \includegraphics{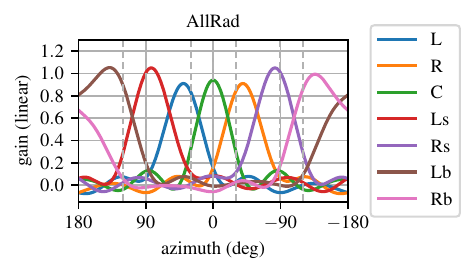}
    \caption{\emph{5OA decoding to 7.0.4}. Loudspeaker gains corresponding to a virtual sound source encoded in 5OA on the horizontal plane at the indicated azimuth. Results with USAT (top) and AllRad (bottom). Only  loudspeakers on the horizontal plane shown.} \label{fig:speaker-gains_decoding_5oa_to_7.0.4}
\end{figure}
\begin{figure}[tbh]
    \centering 
    \includegraphics{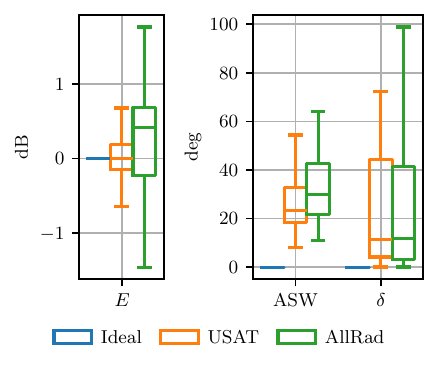}
    \caption{\emph{5OA decoding to 7.0.4}. Box plots indicating the values of the energy in dB ($E$), apparent source width (ASW) and angular error ($\delta$) for the decoding of 5th order Ambisonics to 7.0.4, with USAT (orange) and AllRad (green) methods compared, and ideal values indicated in blue. The boxplots depict the median values, interquartile range, and maximum range (excluding outliers) for a set of directions sampling the upper hemisphere.} \label{fig:boxplot_decoding_5oa_to_7.0.4}
\end{figure}

In this example, USAT is used to decode 5OA into a regular 7.0.4 layout. When used this way, USAT algorithm is essentially equivalent to IDHOA \cite{scaini_ambisonics_2014, scaini_idhoa_2015}, although they differ in some aspects of the implementation (among other things, IDHOA  used a derivative-free algorithm that is much slower than the quasi-newtonian method used in USAT). Results are compared to the well-known AllRad method \cite{zotter_allrad_2012}.

The input configuration for USAT consists of an input matrix ($56\times36$) formed by the gains that encode  a full-sphere t-design cloud of 56  directions sampling the sphere ($L=56$) in 5OA ($M=36$) and an output layout of speakers corresponding to a regular 7.0.4 multichannel ($N=P=11$). We use quadratic cost function coefficients\footnote{It is to be noted that we use a small value for the in-phase coefficient ($c_\Phi^\text{quad}=10$); the goal is not to obtain an in-phase decoding, but rather make the algorithm select positive coefficients whenever possible; without such term, with the other quadratic terms only,  the algorithm has no reason to prefer positive coefficients to negative ones.} specified in Table \ref{fig:table-configurations}, leading to an incoherent or psycho-acoustic decoding.
Regarding AllRad, the same input matrix and output layout are used. The max-rE decoding matrix for the specified layout of speakers is generated with the \emph{AllRADecoder} from the IEM Plug-in suite \cite{iem-plugin}. 

Figure \ref{fig:speaker-gains_decoding_5oa_to_7.0.4} shows that when panning on the horizontal plane, the qualitative behaviour of USAT and AllRad is similar; however, there are subtle differences, such as a more symmetric behaviour of USAT (see Lb, Rb speakers), that will lead to some perceptual differences. These differences become clear in Figure \ref{fig:boxplot_decoding_5oa_to_7.0.4}. On the upper hemisphere USAT outperforms AllRad in two of the three metrics (level and ASW), and the two methods are essentially equivalent on the third (angular error).  Figure \ref{fig:5OA-714-sphere} depicts the 3D reconstruction of those three metrics on the full-sphere, with similar conclusions. USAT presents a quite constant energy distribution, leading to a more homogeneous level perception. Specially remarkable are the generally smaller values of ASW, leading to a more directional Ambisonics decoding with USAT.

The selected USAT parameters result in a smooth decoding, closely aligning with AllRad. However, the versatility of USAT permits alternative configurations to achieve different outcomes, such as more point-like decodings,  albeit at the cost of reduced smoothing in the decoding process. For instance, this can be accomplished by increasing the $c_{IR}$ coefficient, setting a sparsity coefficient, and/or limiting the input matrix to the upper hemisphere.

\subsection{7.0.4 transcoding to 5th order Ambisonics}

\begin{figure}[tbh]
    \centering \includegraphics{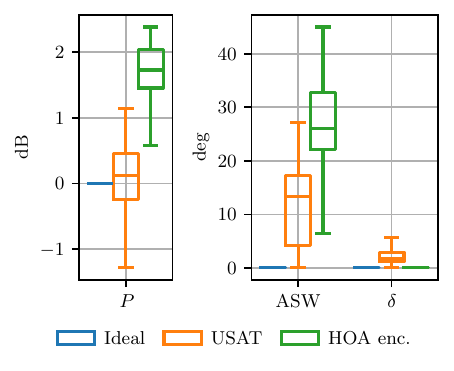}
    \caption{\emph{7.0.4 transcoding to 5OA.} Box plots indicating the values of the pressure, ASW and angular error on a set of points sampling the upper hemisphere. USAT (orange) is compared to a direct encoding of each one of the loudspeakers feeds into 5OA (green). %
    }
    \label{fig:boxplot_transcoding_7.0.4_to_5oa}
\end{figure}

Conversely to the previous case, this example studies the transcoding of a VBAP-encoded 7.0.4 to 5OA.
We evaluate the performance of USAT as an Ambisonics transcoder, in comparison to independently encoding each individual source in Ambisonics format.

For this purpose, the input matrix \Gmat ($54 \times 11$) is formed by the gains needed to encode the set of virtual sources corresponding to the input directions into a 7.0.4 layout ($M=11$) using VBAP. The set of input directions is formed by an upper-half-sphere t-design cloud of 28 points, 15 equi-distant points belonging to the elevation zero plane and 11 points located at the input speakers' positions, with relative weights 6, 3 and 1, respectively (a total of $L=54$). Additionally, to generate the decoding matrix, a set of virtual speakers is provided, formed by a combination of an upper-half-sphere t-design cloud of 30 points and 36 equidistant points belonging to the elevation zero plane (a total of $P=66$). Finally the \Dspk ($66 \times 36$) matrix is the pseudo-inverse matrix that decodes 5OA to the mentioned set of virtual speakers. Using the cost function coefficients specified in Table \ref{fig:table-configurations}, the algorithm delivers an optimized transcoding matrix ($36 \times 11$). The alternative transcoding matrix is  obtained by encoding independently in 5OA each of the 11 loudspeaker feeds at their corresponding directions.

We optimize the pressure and velocity vectors, for compatibility with the physical decoding  matrix \Dspk, which a assumes coherence during the decoding process. These linear decoding coefficients are supplemented with smaller values for the intensity vector coefficients, to provide a clue for the algorithm to continue optimizing in those cases in which the solution is already optimal from the pressure and velocity perspective. Instead, it would also be possible to use a psychoacoustic decoding matrix for \Dspk (e.g.~max-rE; incoherence hypothesis), in which case, for compatibility, we would  use quadratic decoding coefficients. 

Figure \ref{fig:boxplot_transcoding_7.0.4_to_5oa} illustrates the resulting pressure level, ASW and angular error $\delta$.\footnote{In this particular example, the definition of ASW and $\delta$ has been adapted to accommodate for the physical decoding method employed. In this case, we have substituted in equations \eqref{eqs:aswdelta} the intensity vector $\vec I$ with the velocity vector $\vec V$. While experimental validation of ASW and $\delta$ under this revised definition is currently lacking, it aligns with the Gerzon localization principles under ideal decoding conditions.} While the direct encoding does a perfect reconstruction in terms of angular error, the USAT optimized matrix achieves pressure and ASW values  much closer to the ideal. This enhancement is particularly noticeable in the 3D reconstruction presented in Figure \ref{fig:715-5OA-emisphere}.

The increase in the pressure level in the direct Ambisonics encoding method can be attributed to the signal build-up phenomenon: the energy normalization in VBAP (the panning technique used to generate the 7.0.4 input format), is at odds with the linear addition of pressure signals when decoding, leading to an increase in the overall level. USAT is able to detect and correct this signal build-up.

\subsection{5.0.2 decoding to irregular 3.0.1}

\begin{table} [htb]
    \caption{Irregular 3.0.1 layout.} \label{tbl:layout}
    \centering
    \small
    \begin{tabular}{lcc}
        \toprule
        Speaker &  Azimuth  & Elevation  \\
        \midrule
        L & $10\degree$ & $0\degree$ \\
        R & $-45\degree$ & $0\degree$ \\
        S & $180\degree$ & $0\degree$ \\
        T & $0\degree$   &  $80\degree$ \\
            \bottomrule
    \end{tabular}

\end{table}

\begin{figure}[tbh]
    \centering 
    \includegraphics{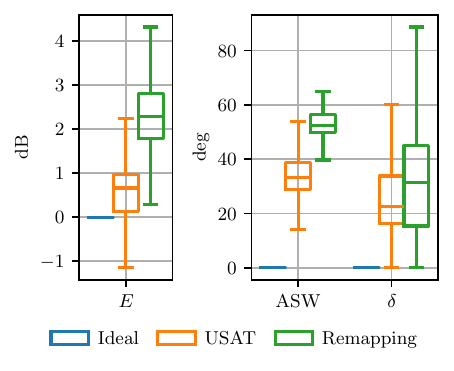}
    \caption{\emph{5.0.2 decoding to irregular 3.0.1}. Box plots indicating the values of the energy, ASW and angular error, for USAT (orange) and channel remapping with VBAP (green).} %
    \label{fig:boxplot_decoding_5.0.2_to_3.0.1_irregular}
\end{figure}

In this instance, USAT generates the decoding matrix of a 5.0.2 format to the irregular 3.0.1 layout detailed in Table \ref{tbl:layout}. Echoing the approach seen earlier, we compare the outcomes to a channel remapping approach, where each input loudspeaker feed is directly decoded into the output format using VBAP.

The set of input virtual sources involves a combination of points, to balance the upper hemisphere behavior, the on-the plane behaviour and the single-channel properties: a t-design cloud of 28 points on the upper-half-sphere, 20 equidistant points on the elevation zero plane, 7 points corresponding to the input speaker positions, and 4 points representing the output layout (with relative weights of 5, 3, 1, and 1, respectively, totalling $L=59$). The output layout consists of 4 loudspeakers ($N=P=4$). Employing the specified cost function coefficients from Table \ref{fig:table-configurations}, the algorithm generates an optimized decoding matrix $4 \times 7$.

Figure \ref{fig:boxplot_decoding_5.0.2_to_3.0.1_irregular} shows that USAT performs  better than the layout remapping using VBAP in all three metrics. USAT is able to correct the significant signal build-up issues present in the remapping method and to improve the directionality of the resulting virtual source, indicated by the significantly smaller ASW and angular error.

\subsection{Audio object decoding to 5.0}
\begin{figure} [bht!]
    \centering 
    \includegraphics{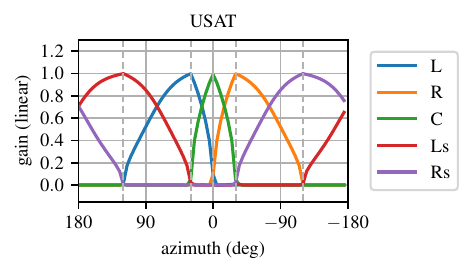} \\
    \includegraphics{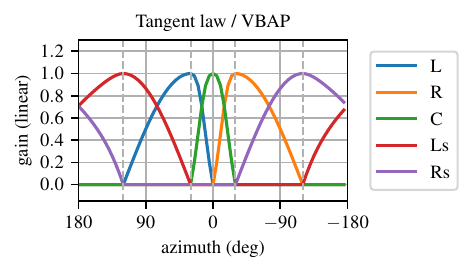} 
    \\
        \includegraphics{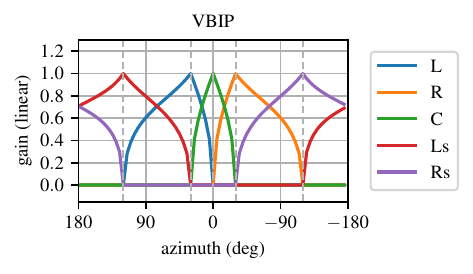} 
    \caption{\emph{Object decoding to 5.0.} 5.0 panning gains for a source at a given position obained from USAT (top), tangent law / VBAP (middle), and VBIP (bottom)} \label{fig:speaker-gains_object_decoding}
\end{figure}

In this final example we illustrate how USAT can be also used as a decoder or transcoder for an audio object-based format, thereby becoming an alternative to a panning law. In particular, we study the decoding to the common 5.0 horizontal layout.

In this case, we optimize for each one of the points in the set of input virtual sources, meaning that the input matrix is an identity matrix of size 72 ($L=M=72$), corresponding to the set of sampled directions on the horizontal plane. The output layout is a regular 5.0 ($N=P=5$). The selected cost function coefficients are shown in Figure \ref{fig:table-configurations}. The optimized decoding matrix given by USAT is $5 \times 72$ (5 gain coefficients for each one of the 72 sampled points).

Figure \ref{fig:speaker-gains_object_decoding} illustrates how USAT panning curves, utilizing the selected optimization coefficients, interpolate between the VBAP panning curves (equivalent to the tangent law) and vector-base intensity panning (VBIP) panning curves \cite{zotter_ambisonics_2019}. While with the chosen optimization coefficients USAT shares similar principles with VBIP, USAT sacrifices some localization accuracy for increased directivity.

\section{Discussion and conclusions} \label{sec:conclusions}

This paper has highlighted the suitability of the USAT algorithm and its associated open-source tool to generate an ideal transcoder or decoder, specifically designed for any input audio format and  any output format or loudspeaker setup. By examining four representative examples, we have shown how USAT often surpasses other state of the art methods across three distinct psychoacoustic metrics: perceived level, apparent source width and angular error.

In particular, it is remarkable how USAT can maximize the directionality, reducing the apparent source width of the rendered virtual sources. It is also noteworthy how USAT automatically corrects the signal build-up issues that often appear when downmixing multichannel mixes to other layouts, without the need of any manual intervention.

It is important to underscore that USAT remains agnostic to the implementation details of both the input and output audio formats. Regarding the input audio format, USAT solely requires information on how a set of virtual audio sources is encoded within the input format. Similarly, for the output format, it only needs knowledge on how to map this specific format to a set of loudspeakers (this mapping being a straightforward 1-to-1 correspondence in the decoding case).
If frequency-dependent decoding matrices are required, the algorithm can be run several times, one per each frequency or frequency band,  each band possibly using different cost function parameters.

In the paper, we have shown generic decoding and transcoding outcomes. However, the features of the transcodings and decodings produced by USAT are customizable by adjusting the cost function parameters, cloud points for evaluation, and relative weights of various spatial zones. Nevertheless, fine-tuning the cost function parameters to achieve desired characteristics often involves a trial-and-error approach: often minor adjustments in the cost function parameters can yield unexpected variations in the results, as it is often the case with optimization problems.

Not only USAT can transcode channel-based formats, USAT can also deal with object-based audio formats. In this sense, USAT is able to provide  the optimal panning laws  to the desired cost function metrics. USAT in general offers two main advantages with respect to conventional panning laws: first, the possibility to adapt to any psychoacoustic target and add custom penalties to the cost function, and second, the ability to address any arbitrary layout in 3D without the need of any additional geometric structure (like a triangulation). A disadvantage of USAT is that finding the optimal panning coefficient with USAT requires solving an optimization problem. In practice, this inconvenience can be addressed by precomputing the panning coefficients on a grid and interpolating over them in real-time.

Lastly, the capability of USAT is restricted to generating fixed linear transcoding matrices, which remain unaffected by the content of the signal to be transcoded. This sets it apart from  signal analysis methods like DirAC \cite{pulkki_dirac_2017} or SASC \cite{goodwin_sasc_2008}, which dynamically adjust the decoding strategy based on incoming signal analysis. USAT's static decoding matrices can be complemented by dynamic signal analysis techniques if neeeded.

\section*{Acknowledgments}

The authors thank Giulio Cengarle and the anonymous reviewers for their useful manuscript feedback.

\bibliography{refs}

\appendix
\section*{Appendix: conventions}

Scalar quantities, including vector and matrix components, are denoted by italic symbols (e.g. \(C_\text{lin}, P_\ell, s_{\ell p}\)).  Matrices of arbitrary dimensions are written in bold symbols (e.g. \Dmat, \Tmat). Unit vectors indicating a direction on the sphere or on the circle are written with a hat on top (e.g. \(\hat u_p, \hat v_l\)). Other vectors in 2D/3D space are written with the arrow symbol on top (e.g. \(\vec V_\ell, \vec I_\ell\)). 

For multichannel layouts, we use a notation $i.j(.k)$, were $i$ is the number of loudspeakers on the horizontal plane, $j$ is the number of low frequency channels (always zero in this paper), and $k$ is the number of overhead loudspeakers.

\begin{figure*}[t]
    \centering
    \includegraphics[width=0.9\columnwidth]{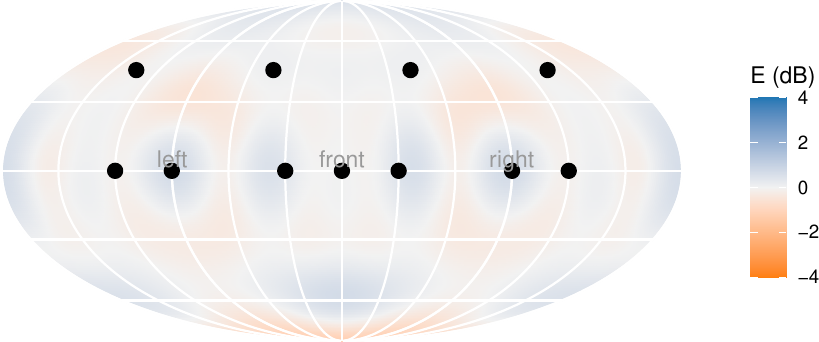}
    \hfill
    \includegraphics[width=0.9\columnwidth]{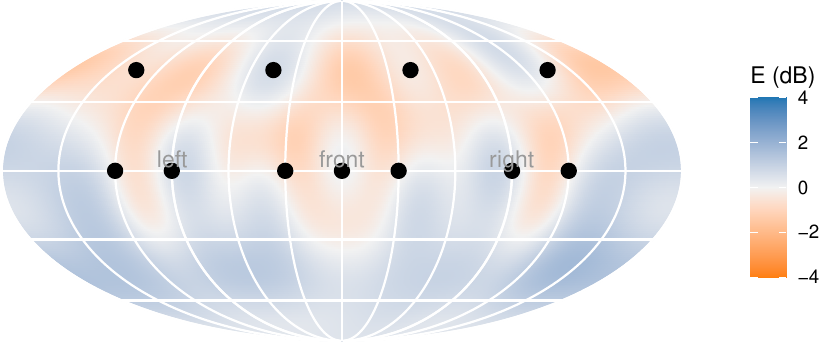}
    \includegraphics[width=0.9\columnwidth]{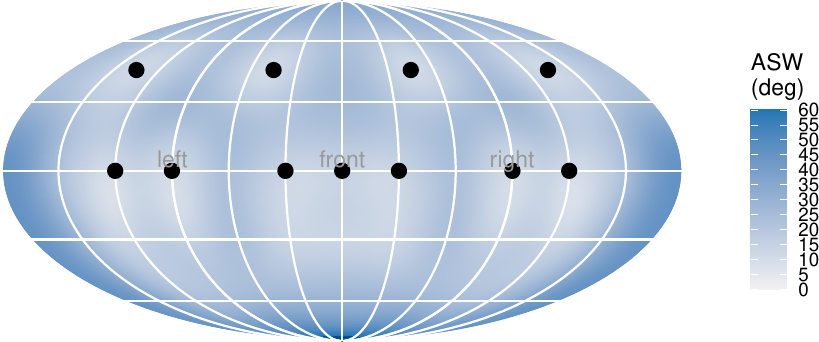}
    \hfill
    \includegraphics[width=0.9\columnwidth]{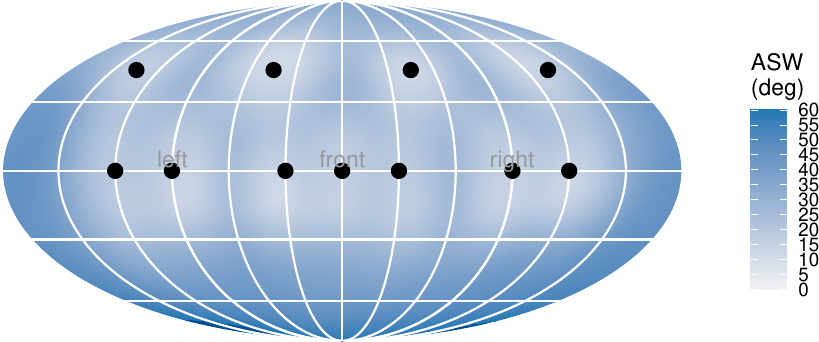}
    \includegraphics[width=0.9\columnwidth]{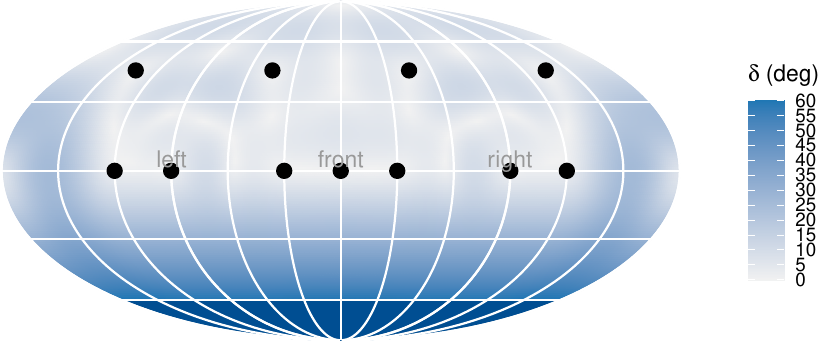}
    \hfill
    \includegraphics[width=0.9\columnwidth]{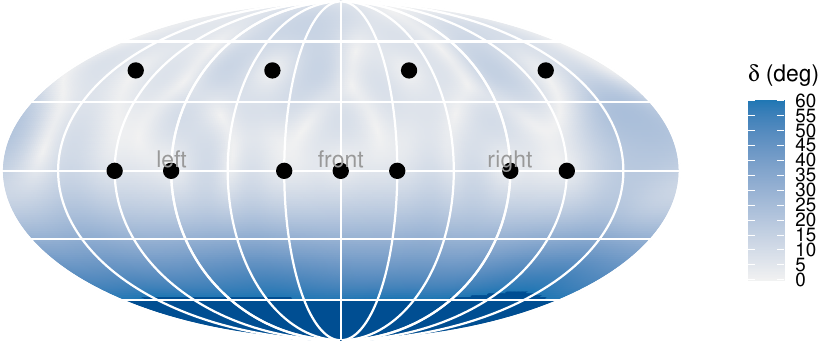}
    \caption{\emph{50A decoding to 7.0.4}. Left column USAT and right column AllRAD. The first row represents the energy reconstruction across the sphere; the second row reports the apparent source width, while the third row the angular error. The black dots represent the 7.0.4 speakers' layout. Values closer to zero (light gray color) indicate better performance.}
    \label{fig:5OA-714-sphere}
\end{figure*}

\begin{figure*}[b]
    \centering
    \includegraphics[width=0.9\columnwidth]{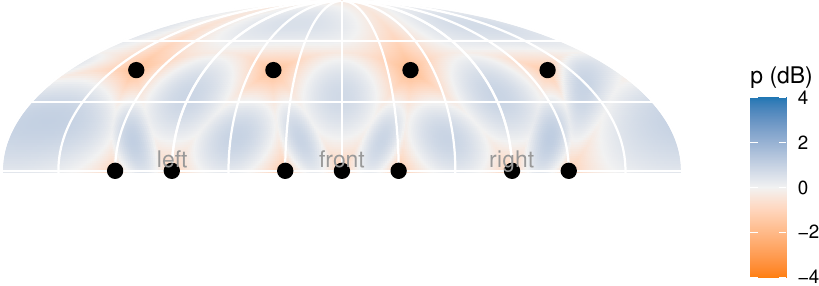}
    \hfill
    \includegraphics[width=0.9\columnwidth]{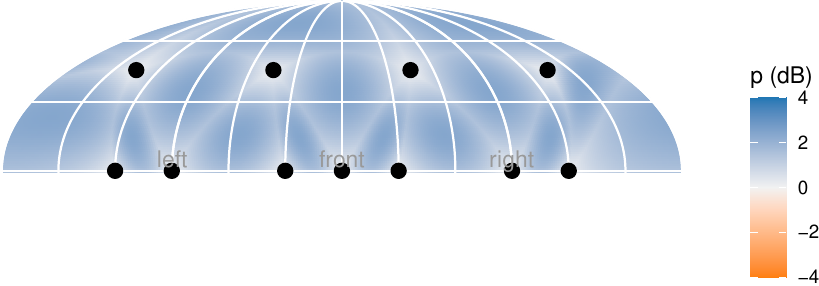}
    \includegraphics[width=0.9\columnwidth]{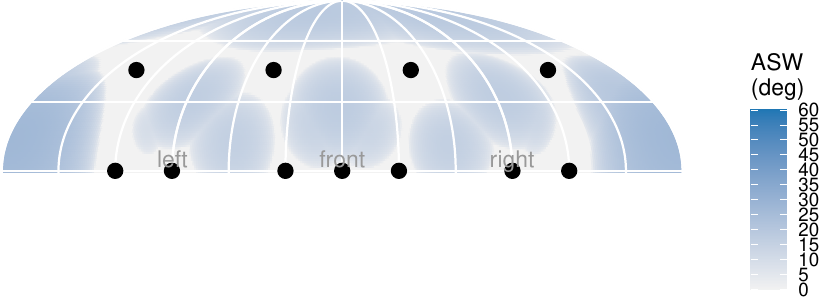}
    \hfill
    \includegraphics[width=0.9\columnwidth]{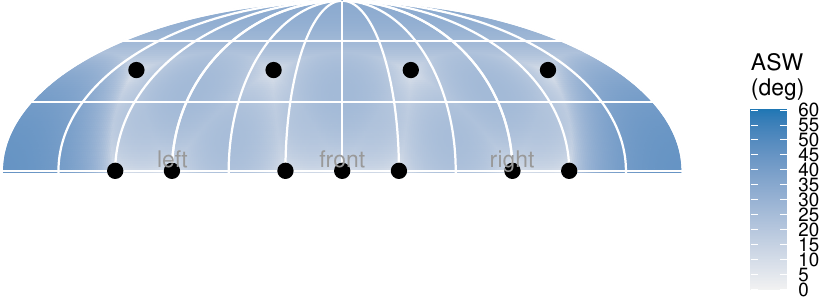}
    \includegraphics[width=0.9\columnwidth]{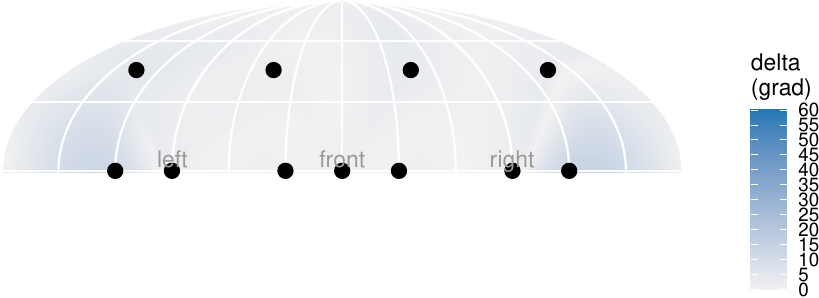}
    \hfill
    \includegraphics[width=0.9\columnwidth]{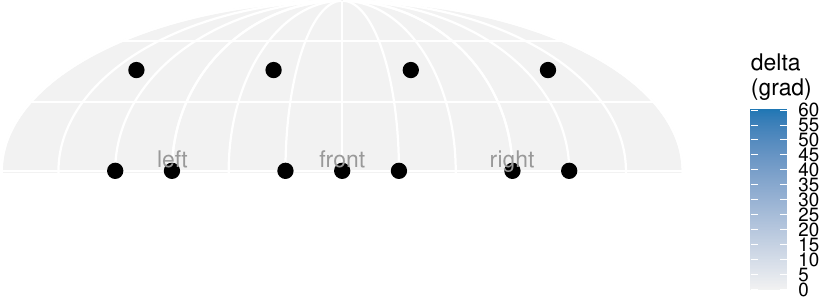}
    \caption{\emph{7.0.4 transcoding to 5OA}. Left column USAT and right column simple source encoding. The first row represents the pressure reconstruction across the sphere; the second row reports the apparent source width, while the third row the angular error. The black dots represent the sources placed in a regular 7.0.4-like configuration. Values closer to zero (light gray color) indicate better performance.}
    \label{fig:715-5OA-emisphere}
\end{figure*}

\end{document}